\documentstyle[12pt]{article}

\newcommand{\adots} {{\mathinner{\mkern2mu\raise1pt\hbox{.}\mkern2mu
\raise4pt\hbox{.}\mkern2mu\raise7pt\hbox{.}\mkern1mu}}}

\newcommand{\np}  {\mbox{\it Nucl. Phys.}}

\newcommand{\pr}  {\mbox{\it Phys. Rev.}}

\newcommand{\frontpg}{
\begin{center}
{\LARGE Renormalization Group at finite temperature in Quantum Mechanics}\\
\vspace*{1cm}
Pierre Gosselin\footnote{Gosselin@ujf-grenoble.fr}$^a$, 
Benoit Grosdidier\footnote{Benoitg@lpli.univ-metz.fr}$^{b}$,
Herv\'e Mohrbach\footnote{Mohrbach@crnvax.in2p3.fr}$^{b}$\\
\vspace*{.5cm}
$^a$Universit\'e Grenoble I, Institut Fourier, UMR 5582 CNRS-UJF, \\
UFR de Math\'ematiques, \\
BP74, 38402 Saint Martin d'H\`eres, Cedex, France\\
\vspace*{.5cm}
$^b$LPLI-Institut de Physique, F-57070 Metz, France\\
\vspace*{2cm}
{\LARGE Abstract}\\
\end{center}
We establish the exact renormalization group equation for the potential 
of a one quantum particle system at finite and zero temperature. As an example we use it 
to compute the ground state energy of the anharmonic oscillator. We comment on
an improvement of the Feynman Kleinert's variational method by the renormalization 
group.
}
\begin{document}
\frontpg
\section{Introduction}
Since most of the path integrals cannot be computed exactly, different 
methods of approximation have been developed. The perturbation expansion is the most 
familiar but it diverges for all couplings strength. An another useful procedure found by 
Kleinert \cite{K} is the convergent variational perturbation expansion.
In field theory the 'exact' Renormalization Group (RG) equations like the Wegner and 
Houghton one \cite{WH} have led to numerous non-perturbative results. In the path 
integral approach a one particle system is similar to a one dimensional spin lattice 
\cite{Roepstorff}, then the RG method could be also useful to extract some non-perturbative 
results for this system. In this paper we derive the RG equation for
the potential of a quantum particle which at finite temperature is not a 
closed expression contrary the field theory case. We show also that  we can 
improve the Feynman-Kleinert's variational method, but cannot compete
with the efficiency of Kleinert's systematic variational perturbation theory.

\section{Renormalization Group Equations in Quantum Mechanics}

Consider the euclidean action of a quantum particle at a finite temperature 

\begin{equation}
S(x)=\int_{0}^{T}\left( {\frac{1}{2}}M\left( {\frac{d}{dt}}x(t)\right) ^{2}
+V(x(t))\right), 
\end{equation}
with $M$ the mass, $V$ the potential and $T=\bar{h}\beta $.
\\
The effective potential is defined as a constrained path integration over periodic 
paths with period $T=\bar{h}\beta $
 
\begin{equation}
\exp \left( -{\beta }V_{0}(x_{0})\right) =\int {\mathcal D} x\delta
(\bar{x}-x_{0})\exp \left( -{\frac{1}{\bar{h}}}S(x)\right).  \label{poteff}
\end{equation}
where $\bar{x}=\frac{1}{T}\int dtx(t)$.
\\
The partition function is 
\begin{equation}
Z=\int {\frac{dx_{0}}{\sqrt{\frac{2\pi \bar{h}T}{M}}}}\exp \left( -{\beta }
V_{0}(x_{0})\right), \label{z}
\end{equation}
We consider the Feynman path integral with
a discretized time $t_{n}={\frac{nT}{N+1}}=n\epsilon $ with $N$ an arbitrary
large number, and $n=0,\ldots ,N+1$. The Fourier decomposition of a
periodic path $x(t_{n})$ contains only a finite number of
Fourier modes
\[
x(t_{n})=x_{0}+{\frac{1}{\sqrt{N+1}}}\sum^{\prime }\exp (i\omega
_{m}t_{n})x_{m}+H.C ,
\]
where $\sum^{\prime }$ is from $1$ to $\frac{N}{2}$ if $N$ is even and from $%
1$ to ${\frac{N-1}{2}}$ if $N$ is odd. The $x_{m}$ are the Fourier modes and 
$\omega _{m}^{2}={\frac{2-2cos{\frac{2\pi m}{N+1}}}{\epsilon ^{2}}}$.
\\
The discretized partition function is then
\[
Z=\int {\frac{dx_{0}}{\sqrt{\frac{2\pi \bar{h}\epsilon }{M}}}}\int \Pi _{1}^{%
\frac{N}{2}}{\frac{dx_{m}d\bar{x}_{m}}{{\frac{2\pi \epsilon \bar{h}}{M}}}}%
\exp \left( -{\frac{1}{\bar{h}}}S_{N}\right), \, 
\]
where the discretized action is given by 
\[
S_{N}(x)=\epsilon \sum_{0}^{\frac{N}{2}}{M}\omega
_{m}^{2}|x_{m}|^{2}-\epsilon \sum_{n=1}^{N+1}V(x(t_{n})).
\]
Using the fact that $\Pi _{1}^{\frac{N}{2}}\epsilon ^{2}\omega _{m}^{2}=\sqrt{N+1}
$ and $T=(N+1)\epsilon $, we define the effective potential as (see \cite{K}): 
\begin{equation}
\exp \left( -{\beta }V_{0}(x_{0})\right) =\int \Pi _{1}^{\frac{N}{2}}{\frac{%
dx_{m}d\bar{x}_{m}}{{\frac{2\pi \epsilon \bar{h}}{\epsilon ^{2}\omega
_{m}^{2}M}}}}\exp \left( -{\frac{1}{\bar{h}}}S_{N}\right) \,.\label{pottef0}
\end{equation}
Instead of evaluating 
(\ref{pottef0}) in one step we integrate mode 
after mode. Let us denote $V_{m}$ the 'running potential'
obtained after ${\frac{N}{2}}-m$ integrations, and $V_{\frac{N}{2}}$ the
initial potential $V$. To find the potential $V_{m-1}$ with
respect to $V_{m}$ we consider paths with only one Fourier mode: 
\[
x(t_{n})=x_{0}+{\frac{1}{\sqrt{N+1}}}exp(i\omega _{m}t_{n})x_{m}+H.C. 
\]
We thus obtain the relation: 
\begin{eqnarray}
&&\exp \left( -{\beta }V_{m-1}(x_{0})\right) \nonumber \\
&&=\int {\frac{dx_{m}d\bar{x}_{m}}{{\frac{2\pi \epsilon \bar{h}}{\epsilon
^{2}\omega _{m}^{2}M}}}}\exp \left( -{\frac{\epsilon }{\hbar }}\left( {M}
\omega _{m}^{2}|x_{m}|^{2}+\sum_{n=0}^{N+1}V_{m}(x_{0}+{\frac{exp(i\omega
_{m}t_{n})x_{m}}{\sqrt{N+1}}}+H.C.)\right) \right) \nonumber \\ 
\label{bvm} 
\end{eqnarray}
Expanding the potential $V_{m}$ around the point $x_{0}$ and summing over
n, yields the following RG equation for the potential

\begin{eqnarray}
V_{m-1}(x_{0})&=&V_{m}(x_{0})+{\frac{1}{\beta }}\log \left( 1+{\frac{
V_{m}^{(2)}(x_{0})}{\omega _{m}^{2}M}}\right) +{\frac{1}{\beta }}\log \left(
1+\sum_{n\ge 2}{\frac{n!}{\beta ^{n-1}}}\;\cdot \right. \nonumber \\
&&\cdot \left. \sum_{k_{1}p_{1}+\ldots +k_{n}p_{n}=n}(-1)^{k_{1}+\ldots +k_{n}}%
{\frac{\left( V_{m}^{(2p_{1})}(x_{0})\right) ^{k_{1}}}{%
k_{1}!(p_{1}!)^{2k_{1}}}}\ldots {\frac{\left( V_{m}^{(2p_{n})}(x_{0})\right)
^{k_{n}}}{k_{n}!(p_{n}!)^{2k_{n}}}}P_{m}^{n}\right) , \nonumber \\
\label{poteff1}
\end{eqnarray}
where all the $p_i\ge 2$ are different and the propagator $P_m={%
\frac{1}{{M}\omega_m^2+{V_m^{(2)}(x_0)}}}$.

\vskip 15pt
Note that in this derivation we have ignored the evolution of 
higher derivative interactions.
The equation (\ref{poteff1}) has not the closed form of the 'exact' WH differential 
equation in field theory \cite{WH} which takes into account only the one loop
contributions resumed in a logarithm term. In field theory the higher order loop
terms are negligible even at finite temperature 
\cite{wetterich} but their influence is taken into account through the running of 
the parameters of the theory. In (\ref{poteff1}) the one loop contributions are
resumed in the first logarithm term but the second one 
contains the higher loop terms. To compare with the usual perturbation expansion 
we can rewrite the RG
equation in the familiar cumulant expansion. Let us define ${\mathcal A}
_{m}=\epsilon \sum_{n=0}^{N+1}\left(
V_{m}(x(t_{n}))-V_{m}(x_{0})-\frac{1}{2}V_{m}^{(2)}(x_{0})x^{2}
(t_{n})\right) $, and
denote the expectation value with respect to the gaussian weight $M\omega
_{m}^{2}+V_{m}^{(2)}(x_{0})$ by $<\ >_{m}$. It is straightforward to derive:
\[
\displaylines{{\frac{1}{l!}}({\frac{-1}{\bar{h}}})^{l}<{\mathcal A}
_{m}^{l}>_{m}=\sum_{n\ge 2}{\frac{n!}{\beta ^{n-1}}}\hfill \hfill \cr
\sum_{k_{1}p_{1}+\ldots +(l-k_{1}\ldots -k_{n-1})p_{n}=n}(-1)^{k_{1}+\ldots
+k_{n}}{\frac{\left( V_{m}^{(2p_{1})}(x_{0})\right) ^{k_{1}}}{
k_{1}!(p_{1}!)^{2k_{1}}}}\ldots {\frac{\left( V_{m}^{(2p_{n})}(x_{0})\right)
^{l-k_{1}\ldots -k_{n-1}}}{k_{n}!(p_{n}!)^{2k_{n}}}}P_{m}^{n}\,.\cr} 
\]
\\
Then defining the cumulants by 
\[
<{\mathcal A}_{m}^{n}>_{m,c}=<\left( {\mathcal A}_{m}-<{\mathcal A}
_{m}>_{m}\right) ^{n}>_{m}, 
\]
the Renormalization Group equation becomes: 
\begin{eqnarray}
V_{m-1}(x_{0})&=&V_{m}(x_{0}) +{\frac{1}{\beta }}\log \left( 1+{\frac{
V_{m}^{(2)}(x_{0})}{\omega _{m}^{2}M}}\right) 
+{\frac{1}{\bar{h}\beta}}<{\mathcal A}_{m}>_m \nonumber \\
&-&{\frac{1}{\beta}}\sum_{l\ge 2}{\frac{1}{l!}}({\frac{-1}{\bar{h}}})^{l}<
{\mathcal A}_{m}^{l}>_{m,c} \label{vmcumul}  
\end{eqnarray}
In this formulation the equation has the disadvantage to mix the powers of 
$\frac{1}{\beta }$.
\\
Clearly the RG equation seems not to be useful to extract nonperturbative
results. Actually it contains an infinite number of terms, negligible
only if $(\frac{V_{m}}{\beta })^p$ are small (for finite temperature the
coupling constants must be small). Nevertheless it is interesting to discuss
the various limiting cases.

\subsection{\protect\smallskip Infinite temperature}

For a large temperature the quantum fluctuations are small and the system
is close to the classical one. The flow of the 
potential stays near the classical one (the potential energy of the
action) and will be obtained after a relatively small number of iterations of
the RG equation.
In particular in the limit  $\beta \to 0$, ${\frac{V_{m}^{(2)}(x_{0})}{%
\omega _{m}^{2}M}}$ and $P$ are of order $\beta ^{2}$ and we obtain
 
\[
V_{m-1}(x_{0})=V_{m}(x_{0})
\]
a running potential constant along the flow. As a consequence the quantum partition 
function reduces to the classical one.

\subsection{Zero temperature}

\smallskip In this opposite case due to quantum fluctuations 
the effective potential is expected to be different from the
classical one and will be obtained after a huge number of iterations of the RG
equation (typically $10^{8}$ for $\beta =10^{5}$). The higher
loop contributions are negligible, then we can extract informations for
large coupling constants. In particular, in the limit $\beta \to \infty $ the
propagator $P_m$ is of order $\frac{1}{V_{m}^{(2)}(x_{0})}$. Then keeping
only the one loop contributions in (\ref{poteff1}) we obtain the closed form:
\begin{equation}
V_{m-1}(x_{0})=V_{m}(x_{0})+{\frac{1}{\beta }}\log \left( 1+{\frac{
V_{m}^{(2)}(x_{0})}{\omega _{m}^{2}M}}\right). \label{vm}
\end{equation}
Recall that for large $N$, $\omega _{m}^{2}$ is equivalent to $\left( {\frac{
2\pi m}{\bar{h}\beta }}\right) ^{2}$ and then runs from $0$ to $\left( {
\frac{\pi }{\epsilon }}\right) ^{2}$. Rewriting  (\ref{vm}) as
\[
V_{m-1}(x_{0})=V_{m}(x_{0})+{\frac{\bar{h}}{2\pi }}{\frac{\pi }{\epsilon }}{
\frac{2}{N+1}}\log \left( 1+{\frac{V_{m}^{(2)}(x_{0})}{\left( {\frac{\pi }{
\epsilon }}{\frac{2m}{N+1}}\right) ^{2}M}}\right)  ,\label{rgwh}
\]
the limit $N\to \infty$ gives a continuous equation
\begin{equation}
V_{k-\Delta k}(x_{0})=V_{k}(x_{0})+{\frac{\bar{h}\Delta k}{2\pi }}\log (1+{
\frac{V_{k}^{(2)}(x_{0})}{Mk^{2}}}), 
\end{equation}
where we have introduced the notations $k^{2}=\omega _{m}^{2}$, $\Delta
k={\frac{2\pi m}{\epsilon (N+1)}}$, and neglected the higher loop
contributions which are of order $(\Delta k)^{2}$. This equation is
the one dimensional Wegner Houghton equation for the effective potential 
\cite{WH}. It is useful to extract informations even for very large coupling
constants as shown below.

\section{Examples}

We solve exactly the Renormalization Group equation for some easy examples
and recover known results.

\subsection{The free particle}

This case is trivial because $V_{\frac{N}{2}}(x_{0})=0$. It's easy to
show recursively that at each step $V_{m-1}(x_{0})=V_{m}(x_{0})=0$. Then,
the effective potential $V_{0}(x_{0})=0$.

\subsection{The harmonic oscillator}

\smallskip As a second example, consider the case: 
$V_{\frac{N}{2}}(x_{0})={\frac{M}{2}}%
\Omega ^{2}x_{0}^{2}$. Once again, we obtain recursively: $%
V_{m-1}(x_{0})={\frac{M}{2}}\Omega ^{2}x_{0}^{2}+C_{m}$ with $C_{m}$ a
constant. Inserting this result in (\ref{poteff1}), we obtain the solution 
\[
V_{0}(x_{0})={\frac{M}{2}}\Omega ^{2}x_{0}^{2}+{\frac{1}{\beta }}\log \left(
\Pi _{1}^{\frac{N}{2}}\left( {\frac{\omega _{m}^{2}+\Omega ^{2}}{\omega
_{m}^{2}}}\right) \right) ,
\]
which gives the partition function (see \cite{K}): 
\[
Z={\frac{1}{\bar{h}\beta \Omega }}\Pi _{1}^{\frac{N}{2}}\left( {\frac{\omega
_{m}^{2}}{\omega _{m}^{2}+\Omega ^{2}}}\right)
\]

\subsection{Perturbation expansion}

Replacing $V_{m}$ by $V_{\frac{N}{2}}$ in (\ref{vmcumul}) 
one recovers the usual perturbative expansion
\begin{eqnarray}
V_{0}(x_{0})&=&V_{\frac{N}{2}}(x_{0})+\sum_{m=1}^{\frac{N}{2}}{
\frac{1}{\beta }}\log \left( 1+{\frac{V_{\frac{N}{2}}^{(2)}(x_{0})}{\omega
_{m}^{2}M}}\right) +\frac{1}{\bar{h}\beta}\sum_{m=1}^{\frac{N}{2}}
<{\mathcal A}_{\frac{N}{2}}>_{m} \nonumber \\
&&-\frac{1}{\beta}\sum_{l\ge 2}\frac{1}{l!}(\frac{-1}{
\bar{h}})^l\sum_{m=1}^{\frac{N}{2}}<{\mathcal A}_{\frac{N}{2}}^{l}>_{m,c} \nonumber
\end{eqnarray}

\section{\protect\smallskip The anharmonic oscillator}

In this part we compute the ground state energy of the anharmonic
oscillator at zero temperature for various values
of the coupling constant by using the RG equation, and compare with Kleinert's \cite{K} 
variational results. The ground state energy is the value of the effective potential 
$V_0(x_{0})$ at its minimum $x_{0}=0$. Unfortunately it is a hard task to use the RG 
equation (\ref{vm}) because at each step we must find the second derivative of the
potential by fitting it numerically. Due to the very slow convergence we must repeat
the procedure a huge number of time for each point of the potential. We
leave this method for an another paper and try instead to compute the flow equations
for the coupling constants. 
\vskip 15pt
We define the n-th coupling constant at the scale $m$ by $g_{m}^{(n)}={\frac{
d^{n}}{dx_{0}^{n}}}_{|{x_{0}=0}}V_{m}(x_{0})$.
\\
We give a
general formulation for the flow of the couplings where we neglect terms of higher
order in $\frac{1}{\beta}$:
\begin{equation}
g_{m-1}^{(0)}=g_{m}^{(0)}+{\frac{1}{\beta }}\log \left( 1+{\frac{
g_{m}^{(2)}}{\omega _{m}^{2}M}}\right). \label{flow1}
\end{equation}
For $k\ne 0$ we have

\begin{equation}
g_{m-1}^{(k)}=g_{m}^{(k)}+{\frac{1}{\beta }}\left(
\sum_{p}(-1)^{p-1}P_{m}(0)^{p}\left( \sum_{\alpha _{1}+\ldots \alpha
_{p}=k, \alpha_i> 0}{\frac{k!A_{p}}{p}}{\frac{g_{m}^{(\alpha _{1}+2)}}{\alpha _{1}!}}
\ldots {\frac{g_{m}^{(\alpha _{p}+2)}}{\alpha _{p}!}}\right) \right) \label{flow},
\end{equation}
where $A_{p}$ is the combinatorial factor of ${\frac{g_{m}^{(\alpha _{1}+2)}
}{\alpha _{1}!}}\ldots {\frac{g_{m}^{(\alpha _{p}+2)}}{\alpha _{p}!}}$ in
the series expansion of $\left( V^{(2)}(x_{0})\right) ^{p}$ in powers of $
x_{0}$ and $P_{m}(0)={\frac{1}{\omega _{m}^{2}M+g_{m}^{(2)}}}$. 

\vskip 15pt

For the initial potential we choose 
\begin{equation}
V_{\frac{N}{2}}(x)=\frac{1}{2}M\Omega^2 x^{2}+\frac{\lambda }{4!}x^{4} 
\end{equation}
The numerical solution of the flow equations (\ref{flow1}) (\ref{flow}) shows a good convergence for 
$N=10^{8}$ and $\beta=10^{5}$. There are no significant improvement for larger 
values of $N$ or $\beta $. At first, we have tried different truncations of the polynomial 
potential and
found the best values by keeping only the couplings $g^{(4)}$ and $g^{(6)}$ as 
is shown in table 1 where we can also read the different values of the couplings 
at the end of the flow when $m=0$. Adding more coupling does not improve the result, on
the contrary. In fact in order to get better results we can not
truncate the effective potential because in quantum mechanics each
coupling constant is relevant. Even when the couplings $g^{(8)}$ and $g^{(10)}$ 
take large negative values they do not influence very much the
flows of the other couplings. Then, it seems that around its minimum the
potential is well fitted by a polynomial interaction of order six. 
\begin{table}
\begin{center}
\begin{tabular}{|l|l|l|l|l|l|}
\hline
$g_{m=0}^{(2)}$ & $g_{m=0}^{(4)}$ & $g_{m=0}^{(6)}$ & $g_{m=0}^{(8)}$ & 
$g_{m=0}^{(10)}$ & $E_{RG}$ \\ \hline
$1.5140$ & $2.4$ & $0$ & $0$ & $0$ & $0.56134$ \\ \hline
$1.4522$ & $1.5343$ & $0$ & $0$ & $0$ & $0.55807$ \\ \hline
$1.4643$ & $1.7706$ & $3.9484$ & $0$ & $0$ & $0.55855$ \\ \hline
$1.4617$ & $1.7110$ & $2.8109$ & $-16.662$ & $0$ & $0.55847$ \\ \hline
$1.4662$ & $1.7236$ & $3.0112$ & $-16.794$ & $-237.46$ & $0.55848$ \\ \hline
\end{tabular}
\vskip 15pt
\caption{comparison of the ground state energy $E_{GR}$ of the anharmonic oscillator
for $M=\Omega=1$, $\lambda=2.4$ and different truncations of the potential. The exact value is 
$E_{exact}=0.55915$ }
\end{center}
\end{table}

A detailed
comparison of the truncation at $x^{6}$ with Kleinert's variational method 
is given in table 2. All the couplings are relevant because they grow 
as we iterate the RG equation. Surprisingly, the
truncation of the potential at the six order gives always
better results than those of the variational method.
The advantage of the latter lies on the fact that one can
improve it systematically with a perturbation expansion which is convergent
(see \cite{K}). In our case the only way to improve it is to consider the
potential as a whole. This requires a numerical study which
is difficult to achieve due to the very slow convergence of the
iterations in the zero temperature case. 

\begin{table}
\begin{center}
\begin{tabular}{|l|l|l|l|l|l|l|}
\hline
$\lambda $ & $E_{exact}$ & $E_{var}$ & $E_{RG}$ & $g_{m=0}^{(2)}$ & 
$g_{m=0}^{(4)}$ & $g_{m=0}^{(6)}$ \\ \hline
$2.4$ & $0.55915$ & $0.5603$ & $0.5585$ & $1.4643$ & $1.7706$ & $3.9484$ \\ 
\hline
$4.8$ & $0.60240$ & $0.6049$ & $0.6014$ & $1.8189$ & $3.2687$ & $12.078$ \\ 
\hline
$7.2$ & $0.63799$ & $0.6416$ & $0.6366$ & $2.2126$ & $4.6951$ & $22.058$ \\ 
\hline
$9.6$ & $0.66877$ & $0.6734$ & $0.6672$ & $2.4041$ & $6.0842$ & $33.296$ \\ 
\hline
$12$ & $0.70618$ & $0.7017$ & $0.6943$ & $2.6614$ & $7.450$ & $45.527$ \\ 
\hline
$14.4$ & $0.72104$ & $0.7273$ & $0.7190$ & $2.9031$ & $8.800$ & $58.589$ \\ 
\hline
$16.8$ & $0.74390$ & $0.7509$ & $0.7417$ & $3.1324$ & $10.137$ & $72.384$ \\ 
\hline
$19.2$ & $0.76514$ & $0.7721$ & $0.7628$ & $3.3514$ & $11.466$ & $86.832$ \\ 
\hline
$21.6$ & $0.78503$ & $0.7932$ & $0.7825$ & $3.5619$ & $12.787$ & $101.87$ \\ 
\hline
$24$ & $0.80377$ & $0.8125$ & $0.8011$ & $3.7648$ & $14.102$ & $117.46$ \\ 
\hline
$240$ & $1.50497$ & $1.5313$ & $1.4982$ & $14.735$ & $128.19$ & $2525$ \\ 
\hline
$1200$ & $2.49971$ & $2.5476$ & $2.4877$ & $41.683$ & $627.76$ & $21462$ \\ 
\hline
$2400$ & $3.13138$ & $3.1924$ & $3.1162$ & $65.742$ & $1250.3$ & $54006$ \\ 
\hline
$12000$ & $5.31989$ & $5.4258$ & $5.2937$ & $190.83$ & $6222.1$ & $460992$
\\ \hline
$24000$ & $6.69422$ & $6.8279$ & $6.6611$ & $302.50$ & $12433$ & $1161228$
\\ \hline
\end{tabular}
\vskip 15pt
\caption{comparison of the exact ground state energy $E_{exact}$ with the variational 
energy $E_{var}$ and the energy $E_{RG}$ obtained from the flow of our truncated potential.}
\end{center}
\end{table}

\subsection{ The double well potential}

In this part we consider the case of the double potential with a frequency $%
\Omega =-1$. The results for some values of the coupling $\lambda $ (see
table 3) show that $g_{m=0}^{(2)}>0$. It means that the quantum
fluctuations smear out completely the double well as it must be, because the
effective potential is a convex quantity at $T=0$. 
For $\lambda =2.4$ our truncation is not accurate enough to get a positive term in the
logarithm of equation (\ref{vm}). For all larger couplings our approximation
is enough to show the spreading of the double potential by the quantum
fluctuations.

\begin{table}
\begin{center}
\begin{tabular}{|l|l|l|l|l|}
\hline
$\lambda $ & $E_{RG}$ & $g_{m=0}^{(2)}$ & $g_{m=0}^{(4)}$ & $g_{m=0}^{(6)}$
\\ \hline
$4.8$ & $0.073$ & $0.366$ & $0.821$ & $4.288$ \\ \hline
$7.2$ & $0.189$ & $0.681$ & $1.955$ & $13.02$ \\ \hline
$9.6$ & $0.266$ & $0.960$ & $3.111$ & $23.34$ \\ \hline
\end{tabular}
\vskip 15pt
\caption{Ground state energy and flow of the coupling constants for the double 
well potential.}
\end{center}
\end{table}

\section{\protect\smallskip The Feynman Kleinert method}

In this section, we give some comments concerning the FK method improved by
the renormalization group. In the FK method, one tries to find a quadratic
potential at each point $x_{0}$ fitting at best the effective potential.
One improves this procedure by looking for a quadratic potential $M\Omega
_{m}x^{2}$ fitting the potential $V_{m}(x_{0})$ at each step of the
renormalization group flow. Then by
improving in such manner the FK method one will take into account some
contributions of the Kleinert's  variational perturbation expansion \cite{K}. 
To achieve this, we follow Kleinert's method
and insert in the right hand side of (\ref{bvm}) the trial
frequency $\Omega _{m}^{2}(x_{0})$ (see \cite{K}).

\begin{eqnarray}
&&\exp\left(-{\frac{T}{\bar h}}V_{m-1}(x_0)\right)
= \int{\frac{dx_m d\bar{x_m}}{{\frac{2\pi\epsilon\bar h}{
\epsilon^2\omega_m^2 M}}}} \exp\left(-{\frac{\epsilon }{\hbar}}M
\left(\omega_m^2+{\Omega^2_m(x_0)}\right)|x_m|^2\right) \nonumber \\
&&\times\exp\left(-{\frac{\epsilon }{\hbar}}\left(\sum_{n=0}^{N+1} 
V_m\left(x_0+{\frac{exp(i\omega_mt_n)x_m}{\sqrt{N+1}}} +H.C.\right)-M
\Omega^2_m(x_0)|x_m|^2\right)\right) \nonumber
\end{eqnarray}
Then, using the Jensen-Peierls inequality 
\[
\int d\mu (x)\exp \left( -O(x)\right) \ge \exp \left( -\int d\mu
(x)O(x)\right) 
\]
for any positive measure $\mu $ normalized to one, and following the same steps as in 
Kleinert's book \cite{K}, we obtain the following RG inequality for the potential
\[
\displaylines{V_{m-1}(x_{0})\le \hfill \cr
\hfill {\frac{1}{\beta }}\log \left( 1+{\frac{\Omega _{m}^{2}(x_{0})}{\omega
_{m}^{2}}}\right) -{\frac{1}{\beta }}{\frac{\Omega _{m}^{2}}{\Omega
_{m}^{2}+\omega _{m}^{2}}}+\int {\frac{dx}{\sqrt{2\pi a_{m}^{2}(x_{0})}}}%
\exp \left( -(x-x_{0})^{2}/2a_{m}^{2}(x_{0})\right) V_{m}(x)\,.\cr}
\]
where
\[ 
a_{m}^{2}(x_0)=\frac{2}{\beta M} \frac{1}{\omega _{m}^{2}+\Omega _{m}^{2}}.
\]
Let us call $V_{a_{m}^{2}}$ the integral on the right hand side of the
inequality. Minimizing the right hand side in the variable $\Omega _{m}^{2}$
we obtain the best approximation for $V_{m-1}$. The frequency
\begin{equation}
\Omega _{m}^{2}(x_{0})={\frac{2}{M}}{\frac{\partial }{\partial a_{m}^{2}}}%
V_{a_{m}^{2}}(x_{0}),
\end{equation}
gives the following equation, which can be solved recursively, starting from 
$\Omega _{m}^{2}(x_{0})=0$ 
\begin{equation}
\Omega _{m}^{2}(x_{0})=\int {\frac{dx}{\sqrt{2\pi }}}\exp \left(
-x^{2}/2\right) {\frac{1}{2\sqrt{a_{m}^{2}}}}V_{m}^{(1)}(x_{0}+\sqrt{
a_{m}^{2}(x_{0})}x).
\end{equation}
Keeping this value  for $\Omega _{m}^{2}$,  enables us to rewrite the
approximation 
\begin{equation}
V_{m-1}(x_{0})={\frac{1}{\beta }}\log \left( 1+{\frac{\Omega _{m}^{2}(x_{0})}{
\omega _{m}^{2}}}\right) -{\frac{1}{\beta }}\frac{\Omega _{m}^{2}(x_{0})}{
\Omega _{m}^{2}(x_{0})+\omega _{m}^{2}}+V_{a_{m}^{2}}(x) \label{rgv},
\end{equation}
which is the variational RG equation. For finite $\beta $, it is possible to
solve (\ref{rgv}), firstly by computing recursively $\Omega
_{m}^{2}(x_{0})$, secondly by calculating $V_{a_{m}^{2}}$ and inserting
this result in the Renormalization group equation. We plan to use this
equation in an another paper. In the following section we check its validity
in the limiting cases.

\subsection{Infinite temperature }

Recall that for large $N$, $\omega _{m}^{2}$ is equivalent to $\left( {\frac{
2\pi m}{\beta }}\right) ^{2}$. Then for $\beta \to 0$, ${\frac{\Omega
_{m}^{2}(x_{0})}{\omega _{m}^{2}}}$ and ${\frac{\Omega _{m}^{2}}{\Omega
_{m}^{2}+\omega _{m}^{2}}}$ are of order $\beta ^{2}$ and  $a_{m}^{2}(x_{0})$
is of order $\beta $.  Inserting these results  in the equation for $
V_{a_{m}^{2}}$ gives $V_{a_{m}^{2}}=V_{m}(x_{0})$.  For $\beta \to 0$ we
recover the usual invariance for the flow of the effective potential: 
\[
V_{m-1}(x_{0})=V_{m}(x_{0})\,.
\]

\subsection{Zero temperature}

In the limit $\beta \to \infty $, $a_{m}^{2}(x_{0})$ is now of order $\frac{1
}{\beta }$. To obtain $\Omega _{m}^{2}$ we expand the potential $V_{m}$ in a
series of $a_{m}^{2}$. To lowest order we obtain: $\Omega
_{m}^{2}(x_{0})={\frac{V_{m}^{(2)}(x_{0})}{M}}$. In the same way, we get 
the lowest order: $V_{a_{m}^{2}}(x_{0})=V_{m}(x_{0})$.
\\
Then keeping only the relevant terms, the limit  for $\beta \to 0$ gives
again the Wegner Houghton equation:

\[
V_{m-1}(x_{0})=V_{m}(x_{0})+{\frac{1}{\beta }}\log \left( 1+{\frac{
V_{m}^{(2)}(x_{0})}{\omega _{m}^{2}M}}\right) 
\]
To  understand why we recover the WH equation, one 
compares with the Kleinert's variational perturbation expansion. In the 
RG equation for the potential at $T=0$, the
contributions of all the higher loops terms are included in the flow of the
couplings. By choosing a  variational frequency which flows in the RG equation
we automatically take into account all the higher loops of the  variational
perturbation expansion and the result must be independent of the
variational frequency. In other words, by integrating mode after mode we
automatically resume the perturbation expansion and we cannot improve it. A
possibility to resume partially the higher loops would be to integrate a
lot of modes at each step, and to deduce the variational frequency for the
resulting potential.  
\section{Conclusion}
The RG equation for the potential of a quantum particle at finite temperature 
was derived. This equation does not allow us to compute non-perturbative 
quantities. Then we suggest to use the Feynman Kleinert's variational method 
improved by the RG. At zero temperature we recovered the Wegner Houghton equation 
which was used to compute the ground state energy of the anharmonic oscillator.
It would be also interesting to compute the RG equation for the effective action. 
Our preliminary work show the generation of non local interactions. We also plan 
to apply the RG procedure for systems with many quantum particles in the 
discretized path integral representation.

\end{document}